**Weber Lazar MOM Histograms 20190902 .pdf – 8,372 words.**

# Method of Moments Histograms

James S. Weber
Nicole A. Lazar
September 2, 2019

**Abstract**. ***Uniform bin width histograms are widely used so this data graphic should represent data as correctly as possible. Method of moments based on familiar mean, variance and Fisher-Pearson skewness cure this problem.***

## 1. Introduction.

There are several motivations for this work. Mainly we challenge the belief that widely used uniform bin width histograms reliably display data skewness, variance, and mean, and can suggest normal or non-normal data distributions. However, histograms often are not visually consistent with the first three standardized moments. Method of moments (MOM) histograms correct this problem, is more intuitive, pragmatic than other definitions of "good" histograms (and density estimators) such as MISE - Rudemo (1982), Silverman (1986), Scott (1992); shape stability - Simonoff & Udina (1997); histosplines – Minnotte (1998), etc. (Further, Minnotte, 1998, p 667 observes "Smaller datasets, ...will be better served by a simpler density estimate, such as a histogram ...")

Although K. Pearson (circa 1890) advocated both histograms and MOM, he did not include histogram densities in his families of distribution curves and did not develop MOM histogram densities. (Elderton, 1906; Elderton, Johnson, 1969; Pearson, 1894, 1895)

Our computational procedure involves histogram shape level sets (to be defined shortly) that lead to all histogram shapes, transparency for MOM histograms, and a framework for comparing many kinds of "good" histograms. This is novel and may extend to level sets for other data aggregations. Even though our primary focus is MOM histograms, there is interest in the tension between elementary pragmatic methods, and sophisticated approaches (see Little, 2013, Hoaglin, Mosteller, Tukey, 2000); Pearson's advocacy of histograms and MOM (1894, 1895); obtaining all histogram shapes that data can have including paradoxical shape reversals; exact implementation of MISE, Simonoff & Udina shape stability criteria; and comparing histograms according to various optimality criteria, especially for smaller samples.

## 1.1 A Skewness Reversal Example.

Histogram skewness provides the most conspicuous anomaly even though mean and variance can be troublesome, Cooper & Shore, (2008), Doane, Seward (2011). To emphasize that histograms do not always reflect data skewness and can even show skewness opposite of data skewness, Figure 1 shows two histograms, at least one of which misrepresents skewness. And both could be wrong!

Financial ratio data was examined with EXCEL. Histogram A has default location anchor point .9355 and bin-width .0326. Four decimal places were considered inappropriate, so .9355 was trimmed to .93, .0326 to .03, uniform width bins and bin counts were recalculated leading to Histogram B. This should not have changed shape much, if at all. However, Histogram B shows shape and skewness reversal.

**Figure 1 – Two views of financial ratio data, Data #1**

| *EXCEL Bins* | |
|---|---|
| Bin | Frequency |
| 0.9355 | 0 |
| 0.9681 | 1 |
| 1.0007 | 5 |
| 1.0333 | 9 |
| 1.0659 | 12 |
| 1.0985 | 1 |
| More | 2 |

| *Trimmed Bins* | |
|---|---|
| Bin | Frequency |
| 0.96 | 0 |
| 0.99 | 2 |
| 1.02 | 12 |
| 1.05 | 9 |
| 1.08 | 4 |
| 1.11 | 2 |
| More | 1 |

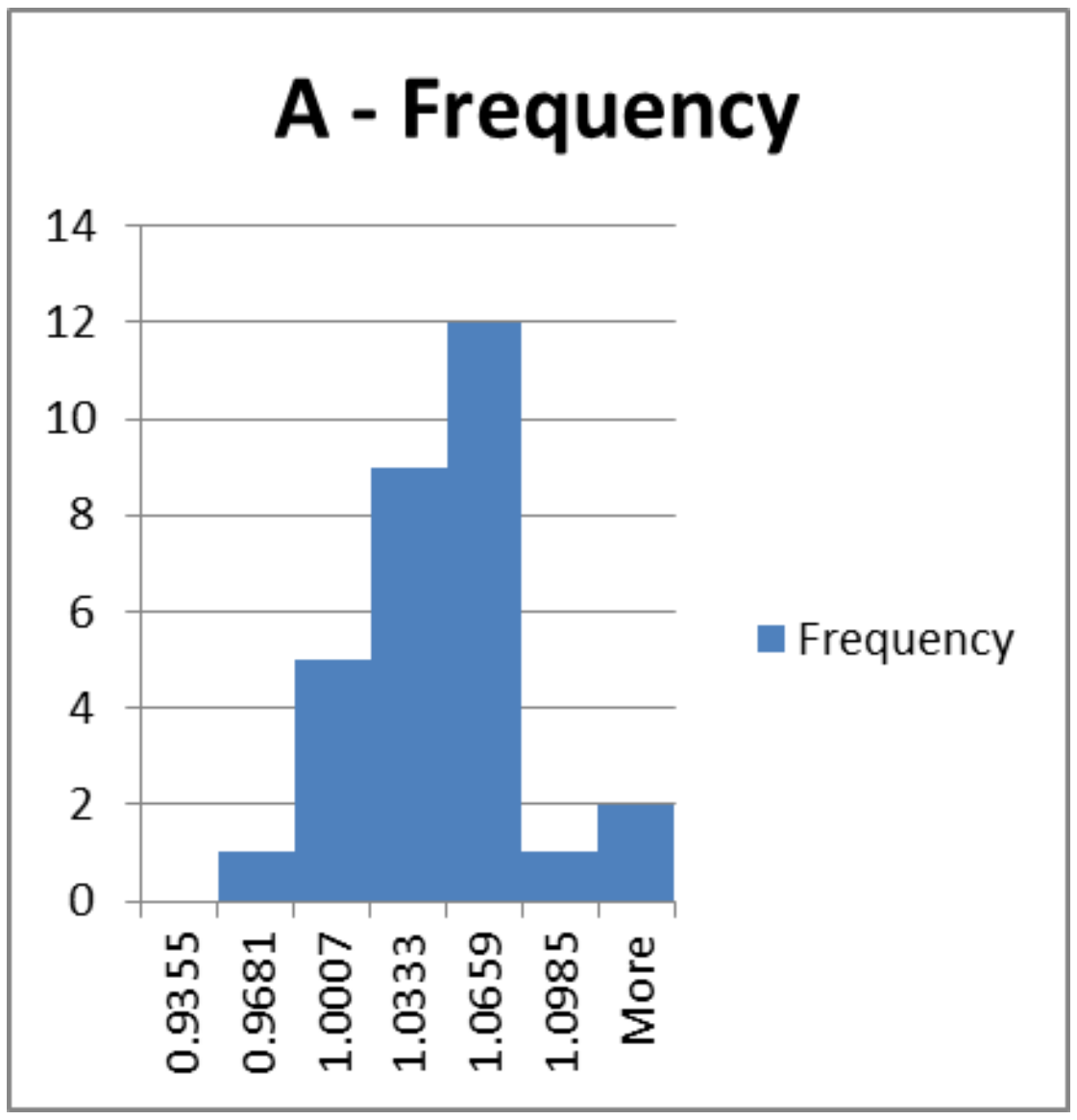

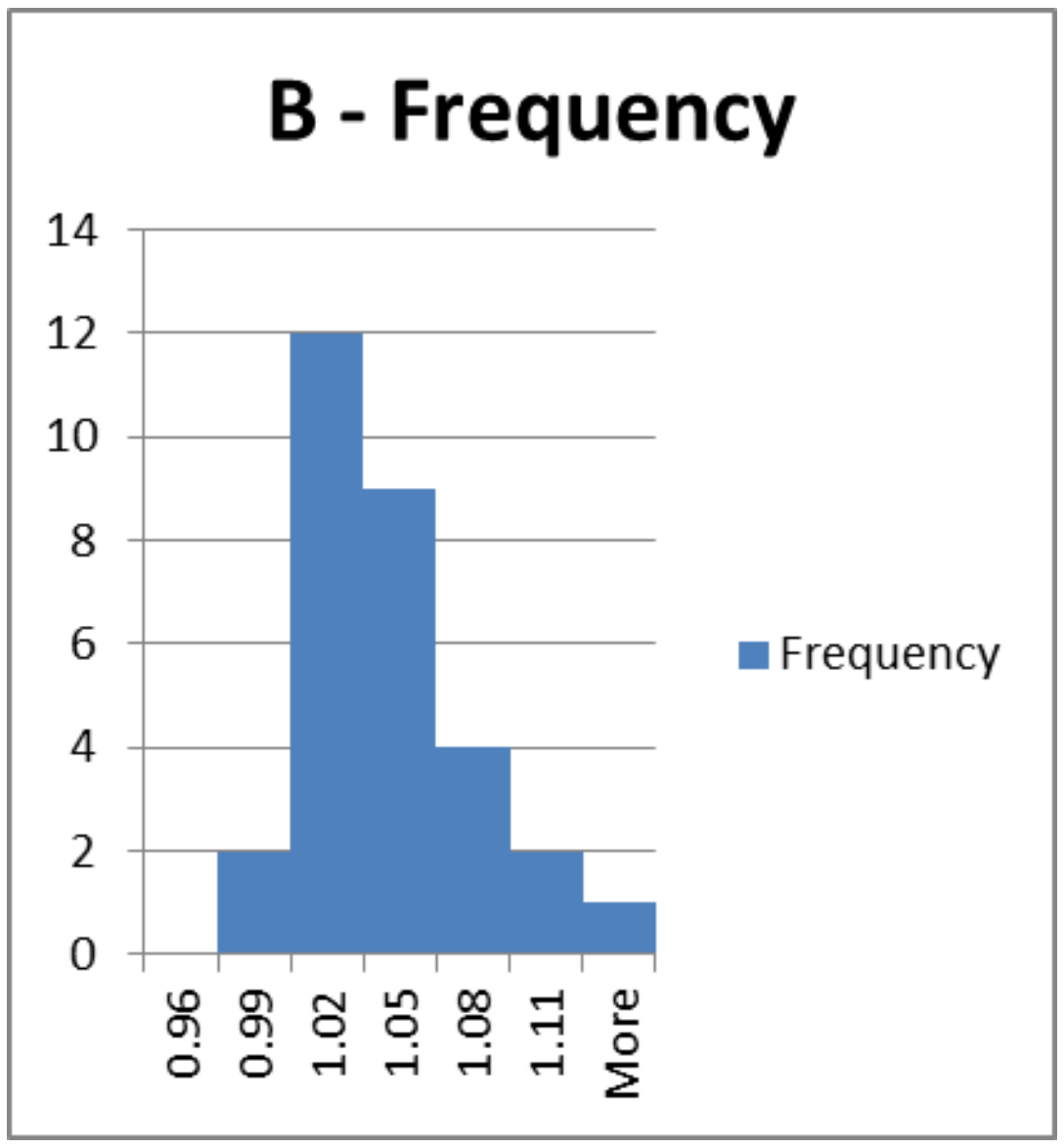

| | | | | | | | | | | | |
|---|---|---|---|---|---|---|---|---|---|---|---|
| 1 | **0.968** | 6 | **0.999** | 11 | **1.012** | 16 | **1.035** | 21 | **1.042** | 26 | **1.056** |
| 2 | **0.982** | 7 | **1.004** | 12 | **1.015** | 17 | **1.037** | 22 | **1.042** | 27 | **1.059** |
| 3 | **0.991** | 8 | **1.004** | 13 | **1.017** | 18 | **1.037** | 23 | **1.047** | 28 | **1.081** |
| *4* | **0.993** | *9* | **1.007** | *14* | **1.019** | *19* | **1.039** | *24* | **1.053** | *29* | **1.107** |
| *5* | **0.998** | *10* | **1.01** | *15* | **1.021** | *20* | **1.039** | *25* | **1.055** | *30* | **1.131** |

If both Histograms A and B appeared during exploratory or instructional discussions, or a data science workshop, could a professor, instructor, scientist, or consultant explain how uniform bin width histograms can portray data as both positively and negatively skewed? Based on interaction with colleagues as well as comments in Huff (1954), Minieka & Kurzeja (2000), Moore et al (2002, 2009), Ramsey (2001), *we believe the answer is "No."*

## 2 Method of Moments Frequency and Density Histograms

Uniform bin width frequency, relative frequency, and density histograms are graphs associated with bin counts, $v_k$, $k$ =1 to $\boldsymbol{K}$, for data, $x_i$, $i$ = 1 to $n$ and half-open bins $[a, b)$ (Kendall, Stewart, 1963, among others). Bins are written $[t_o + (k\text{-}1)h, t_o + kh)$ with parameters $t_o$, $h$ for bin location anchor point and bin width, $t_o \le x_{min} < t_o + h$. $k$ is a bin index, $k$ =1 to $\boldsymbol{K}$, wherein $\boldsymbol{K}$ is the maximum number of bins. Bin heights are *proportional* to bin counts. Vertical scales distinguish frequencies, relative frequencies, and histogram density values however the overall appearance of these histograms is the same.

### 2.1 Histogram Shape, MOM Mean, and Variance Constraints.

Histogram *shape* usually refers to the *list* of bin counts, $v_k$. We focus on shapes, data, and consistency of shape with data mean, variance, and skewness, in contrast to continuously varying bin width and location to asymptotically optimize criteria for histogram densities and equivalent frequency histograms.

Histogram shape level sets, described shortly, enable transparent understanding and calculation of MOM frequency and density histograms, as well as insight into other "optimal" histograms such as MISE (Scott, 1992), optimal bin width determined via shape stability (Simonoff & Udina, 1997), and the variability of shape associated with number-of-bins and bin-width rules. (We focus on frequency histograms because they are more familiar and widely used than density histograms. Also, MOM frequency and density histograms are essentially identical. (Frequency and density histogram mean and skewness constraints are identical. Variance constraints differ by $h^2/12$.)

A text-book example of MOM is estimation of normal distribution parameters μ and $\sigma^2$, with $\bar{x}$ and $s_x^2$ (e.g. Lindgren, 1968 p 280, 5-16, p 507, 5-16.) However, histogram bin parameters $t_o$, $h$ are not distribution parameters like μ, σ, and not directly associated with data features such as μ, σ, and skewness.

### 2.2 Computation of Mean and Variance MOM histograms.

It is instructive to begin MOM histograms in a familiar way. MOM usually employs lower moments to define parameters. So even though incorrect skewness is the most conspicuous, we first consider calculating $t_o$, $h$ from mean and variance constraints (Lindgren, 1968 p 279, 5-12, p 280 5-16 p 507, 5-16). Achieving mean and variance (MV) consistency with the data for bin counts $v_k$ begins with a histogram shape and calculating new bin parameters, ($t_o$, $h$), from variance and mean constraints, (1), (2ab) below, using bin midpoints and bin frequencies. (Weber 2005-2016 inclusive, esp. 2016b)

Variance constraint (1) depends only on bin width, $h$, so first obtain MOM bin width, $h^{\text{mom}}$, for shape $v_k$, from (1). Then calculate MOM bin location, $t_o^{\text{mom}}$ from $h^{\text{mom}}$ and (2a) or (2b). Sample size is $n$, sample mean is $\bar{x}_x$, grouped data mean based on bin midpoints and bin frequencies is $\bar{x}_g$, relative frequency weighted average bin index is $\bar{k}$ $= \frac{1}{n} \sum_{k=1}^{K} v_k k$, sample variance is $s^2_{\text{x}}$, and grouped data variance is $s^2_{\text{g}}$.

**Frequency histogram MOM variance constraint leads to MOM bin width, $h^{mom}$**

*grouped data* variance = *sample* variance

$$s^2_{\text{g}} = s^2_{\text{x}}$$

$$s^2_{\text{g}} = \frac{h^2}{n-1}\left[\sum_{k=1}^{K} v_k (k-\bar{k})^2\right] = s^2_{\text{x}}$$

$$\boldsymbol{h^{mom}} = s_{\text{X}} \left[(n-1)/ \sum_{k=1}^{K} v_k (k-\bar{k})^2\right]^{½} + \mathbf{0}\, t_{\text{o}} \qquad (1)$$

**Frequency histogram MOM mean constraint $h^{mom}$ leads to $t_0^{\text{mom}}$**

*grouped data* mean, $\bar{x}_g$ = *sample* mean, $\bar{x}$

$$\bar{x}_g = \frac{1}{n} \sum_{k=1}^{K} v_k [t_0 + (k-\frac{1}{2})h] = t_o - \frac{h}{2} + \frac{h}{n}\sum_{k=1}^{K} k v_k = t_0 + h\,(\bar{k} - \frac{1}{2}) = \bar{x}_x$$

$$\boldsymbol{t_0}^{\text{mom}} = \bar{x}_x - \boldsymbol{h}^{\text{mom}} (\bar{k} - \frac{1}{2}) \text{ and } \boldsymbol{h}^{\text{mom}} \text{ from (1), leads to} \qquad (2a)$$

$$\boldsymbol{t_0}^{\text{mom}} = \bar{x}_x - s_{\text{X}} \left[(n-1) / \sum_{k=1}^{K} v_k (k-\bar{k})^2\right]^{½} (\bar{k} - \frac{1}{2}) \qquad (2b)$$

Once ($\boldsymbol{t_0}^{\text{mom}}$*,* $\boldsymbol{h}^{\text{mom}}$) have been obtained from (1), and (2), bin counts, $\boldsymbol{v_{\text{k}}}^{\text{mom}}$, must be recalculated with MOM revised bins:

$$[\boldsymbol{t_0}^{\text{mom}} + (k-1)\, \boldsymbol{h}^{\text{mom}}, \boldsymbol{t_0}^{\text{mom}} + k\boldsymbol{h}^{\text{mom}}),\ k = 1, \ldots \boldsymbol{K}.$$

*If recalculated bin counts are the same*, i.e. $\boldsymbol{v_{\text{k}}}^{\text{mom}} = \boldsymbol{v_{\text{k}}}$, *then shape*, $\boldsymbol{v_{\text{k}}}$ *is* MV *jointly consistent with the data mean and variance. If not, shape* $\boldsymbol{v_{\text{k}}}$ is *not* MV *jointly consistent with the data mean and variance*.

Also, joint MV consistency does not identify a unique MOM uniform bin width histogram. Table 1 shows three MV jointly consistent shapes: **(3,4,5)**, **(5,3,4)**, (**1,2,3,1,2,3**); and three not MV jointly consistent shapes: (**5,4,3**), (**1,2,3,3,2,1)** (**3,2,1,1,2,3**), for Data #3, Weber (2008a). So MV *jointly consistent frequency histogram shapes are not unique and are not similar*. Although "similar" is subjective, to the authors, shapes (**3,4,5**), (**5,3,4**), (**1,2,3,1,2,3**) are not similar. We will use skewness to identify preferred MV consistent shapes.

**Table 1**

| | **Shape** | $t_o$ | $h$ |
|---|---|---|---|
| **Shape #** | **Jointly** | **Consistent** | |
| **1 -** $v_k$ | **(3,4,5)** | 0.3690 | 1.8130 |
| $v_k^{mom}$ | **(3,4,5)** | -0.4602 | 2.1981 |
| **2 -** $v_k$ | **(5,3,4)** | 0.3250 | 2.0480 |
| $v_k^{mom}$ | **(5,3,4)** | 0.3159 | 2.0382 |
| **3 -** $v_k$ | **(1,2,3,1,2,3)** | **?** | **?** |
| $v_k^{mom}$ | **(1,2,3,1,2,3)** | -0.2931 | 1.0489 |
| | **NOT Jointly** | **Consistent** | |
| **4 -** $v_k$ | **(5,4,3)** | 0.3159 | 2.1282 |
| $v_k^{mom}$ | **(3,3,1,4,1)** | 2.7459 | 1.3724 |
| **5 -** $v_k$ | **(1,2,3,3,2,1)** | 0.0090 | 1.1200 |
| $v_k^{mom}$ | **(1,2,3,3,0)** | -0.6039 | 1.2691 |
| **6 -** $v_k$ | **(3,2,1,1,2,3)** | 0.3590 | 0.9990 |
| $v_k^{mom}$ | **(1,2,1,2,1,2,3)** | 0.5400 | 0.8878 |

$n = 12$

| $i$ | $x_i$ |
|---|---|
| 1 | **0.37** |
| 2 | **1.13** |
| 3 | **1.23** |
| 4 | **2.25** |
| 5 | **2.35** |
| 6 | **2.45** |
| 7 | **3.37** |
| 8 | **4.37** |
| 9 | **4.47** |
| 10 | **5.37** |
| 11 | **5.47** |
| 12 | **5.61** |

Non-unique shape for histograms that are consistent with data mean and variance shows that histogram skewness is not the only problem! The reader may question how two apparently non-dependent linear constraints, (1), (2a) can lead to many MOM MV consistent shapes? Constraints (1), (2a) depend on bin counts $v_k = v_k(x_i;\, t_o,\, h)$ so that constraints (1), (2a) are *piecewise* linear in $t_o$, $h$ for sets of $t_o$, $h$ values for which the bin counts are the same, i.e. for ($t_o$, $h$) in *shape level sets*, described in the next section. Different shapes lead to different linear constraints (1), (2a) and different calculated MOM $t_o$, $h$ values. Shapes can be MV jointly consistent or MV not jointly consistent.

Searching for MOM histograms is about four constraints for shape, mean, variance and skewness. Sample data, $x_i$, are constants and bin parameters, $t_o$, $h$. are variables. Bin counts, $v_k$, are a function of $x_i$, $t_o$, $h$. We need to know possibilities for $t_o$, $h$ for uniform width bins so that grouped data mean, variance and skewness, or histogram density mean, variance and skewness agree as closely as possible to data sample mean, variance and skewness. For fixed $x_i$, bin counts, $v_k$, are shape constraints in that not all lists of integers that add up to the sample size can occur as uniform bin width histogram bin counts for $x_i$. If a shape is possible, Table 1 shows that mean and variance constraints cannot always be satisfied, and shapes associated with $t_o$, $h$ values that satisfy the mean and variance constraints are not similar. Graphs in the two dimensional $\{(t_o, h)\}$ plane clarify consistency of histogram shape level sets with sample mean, variance constraints.

### 2.3 Illustration and Computation of Shape Level Sets, then Shapes.

**Definition**: For data $x_i$, *uniform bin width histogram shape level sets are convex polygons of* $(t_o, h)$ values leading to the same shape for half-open bins $[t_o + (k-1)h,\ t_o + kh)$, $k = 1$ to $\boldsymbol{K}$.

Fig 2 shows uniform width bin histogram shape level sets for {1, 2, 5} for at most four bins. (This *illustrative* data is employed simply because the number of shapes and shape level sets grows rapidly. Many data values, shape level sets and shapes quickly obscure the picture.)

*Shape level set vertices are obtained from intersections of straight-line boundaries*, (3b), below, within a $(t_o, h)$ bounded domain, $\boldsymbol{D}_o$, defined primarily so that a first bin, $[t_o, t_o + h)$, contains the data minimum.

To clarify (3b), changes in $t_o$ and $h$ lead to changes in bin edges $t_o + kh$. If a bin edge moves past a data point, $x_i$, then $x_i$ is counted in an adjacent bin, leading to a different shape. So, shape level set boundaries in $\{(t_o, h)\}$ are lines (3a) and (3b).

$$\text{bin edge} = \text{data value} \qquad (3a)$$

$$t_o + kh = x_i,\ \ k = 1 \text{ to } \boldsymbol{K},\ i = 1 \text{ to } n \qquad (3b)$$

**Figure 2 Shapes and Shape Level sets in {($t_o$, $h$)}**

(**3**), Light Orange
(**1, 2**), Purple
(**2, 1**), Red
(**1, 1, 1**), Black
(**2, 0, 1**), Green
(**1, 1, 0, 1**, Dark Orange
(**2, 0, 0, 1**), Light Blue

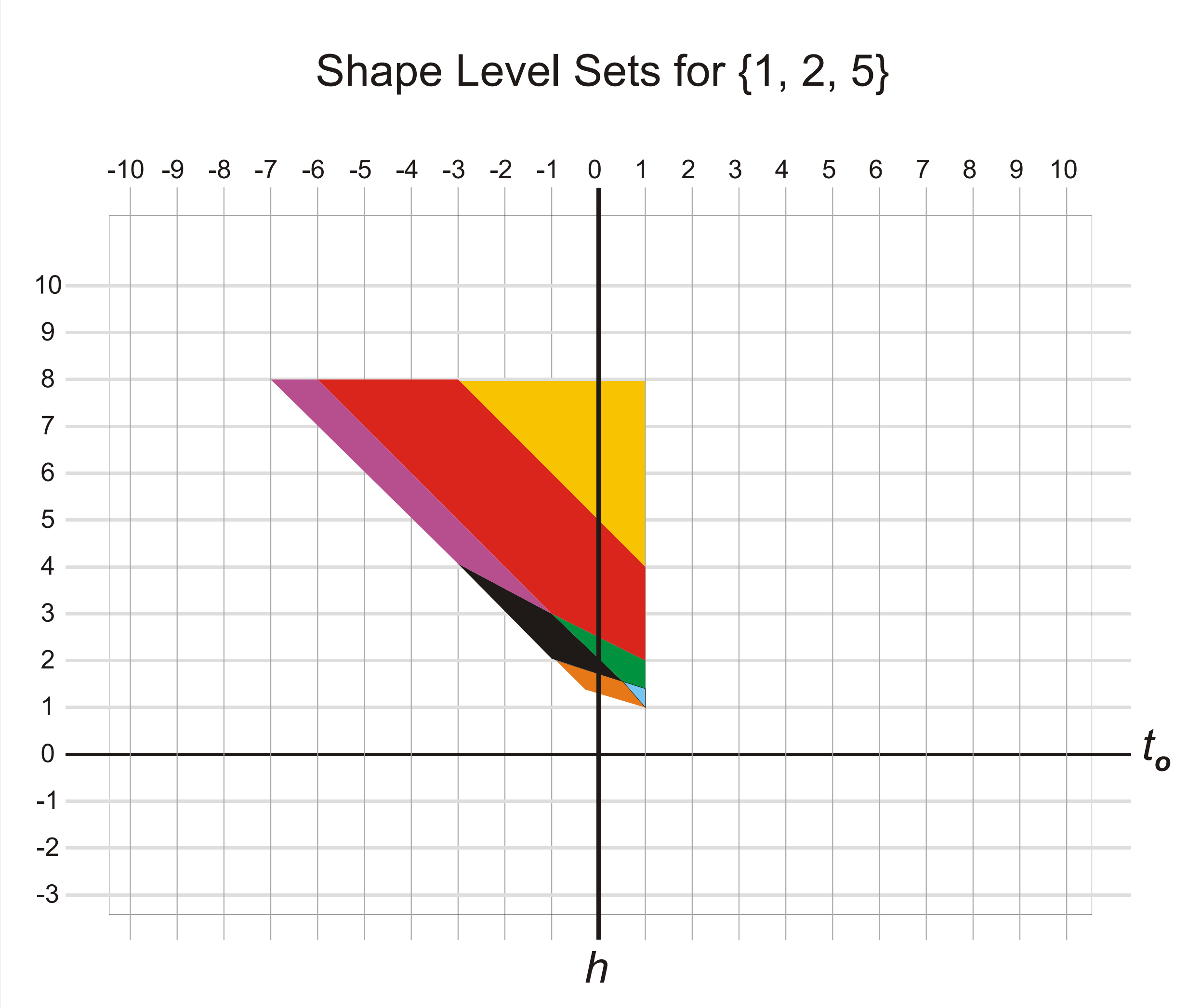

Each polygon interior corresponds to $t_o$, $h$ values leading to uniform width histogram bins that lead to the same bin counts for data {1, 2, 5}. In linear programming and economics, “shape level sets” are “feasible sets.” Lindgren (1968, pp 18, 362,7-8) mentions “These partition sets are ‘level curves’ or ‘level surfaces’ …” However, shape level sets are not curves or surfaces. The concept is pervasive, but apparently not previously applied to histogram shapes. Shape level sets also might be called "inverse images." Bin edge discontinuity leads to level set boundaries belonging to *one or the other* of adjacent level sets.

We show explicit details for calculating $\boldsymbol{D}_o$ vertices and lines (3b) that partition $\boldsymbol{D}_o$ into shape level sets. The simplicity and rationale for (3b) is compelling, however it is "overkill" in that not every line is "effective," "binding," or "active." The simplest example of non-binding, non-effective, or non-active constraints arises from the "At most $\boldsymbol{K}$ bins." This constraint with right open bins, $[t_o + (k-1)h, t_o + kh)$, and $t_o \le x_{(1)}$ leads to the inequalities $x_i < t_o + \boldsymbol{K}h$, i.e. $x_{(1)} < t_o + \boldsymbol{K}h$, $x_{(2)} < t_o + \boldsymbol{K}h$, … $x_{i(n)} < t_o + \boldsymbol{K}h$. Because $x_{(1)} < x_{(2)} <$ … $x_{i(n)}$, the constraints $x_i < t_o + \boldsymbol{K}h$, i.e. $x_{(1)} < t_o + \boldsymbol{K}h$, $x_{(2)} < t_o + \boldsymbol{K}h$, … $x_{(n-1)} < t_o + \boldsymbol{K}h$ are all implied by the last one, $x_{i(n)} < t_o + \boldsymbol{K}h$, and not binding, effective, active. Similarly, (3b) leads to many lines and associated inequalities that are outside of $\boldsymbol{D}_o$ and not a boundary of a shape level set inside of $\boldsymbol{D}_o$. Usually it is not necessary to indicate which are binding, which are not, except here for clarity.

**A.** $x_{min}$ in first bin, $k = 1$, for bins $[t_o + (k-1)h, t_o + kh)$ → $t_o \le x_{min} < t_o + h$

**B.** For at most $\boldsymbol{K}$ bins → $t_o \le x_{max} < t_o + \boldsymbol{K}h$

(For exactly $\boldsymbol{K}$ bins to satisfy number-of-bins rules: $t_o + (\boldsymbol{K}-1)h \le x_{max} < t_o + \boldsymbol{K}h$ )

**C.** To bound $\boldsymbol{D}_o$: $h \le x_{max} - x_{min} + \Delta$, $0 < \Delta$. We choose $\Delta = x_{max} - x_{min}$, so $h \le 2(x_{max} - x_{min})$

**A**. $t_o \le 1 < t_o + h$ → $t_o = 1$, $t_o + h = 1$
**B**. $t_o \le 5 < t_o + 4h$ → $t_o = 5$, $t_o + 4h = 5$ (But $t_o = 5$ is not a binding constraint.)
**C**. $h \le 2(x_{max} - x_{min})$ → $h = 8$

Effective $\boldsymbol{D}_o$ boundaries:
**i**. $t_o = 1$
**ii**. $t_o + h = 1$
**iii**. $t_o + 4h = 5$
**iv**. $h = 8$

$\boldsymbol{D}_o$ vertices:
vertex$_1$: (1.00, 1.00), **i**. & **iii**.
vertex$_2$: (1.00, 8.00), **i**. & **iv**.
vertex$_3$: (–7.00, 8.00), **ii**. & **iv**.
vertex$_4$: (–0.33, 1.33), **ii**. & **iii**.

**Shape Level Set – SLS - boundaries**

bin edge = data value, (3a)

$t_o + kh = x_i$, $i = 1$ to $n$ for $n = 3$; $k = 0$ to $\boldsymbol{K}$ for $\boldsymbol{K} = 4$; leading to:

**(3b) - *k* = 0**

$t_o + 0h =$ 1, same as **i**. above
$t_o + 0h =$ 2, not binding
$t_o + 0h =$ 5, not binding

**(3b) - *k* = 1, 2, 3, 4**

**1.** $t_o + h = 1$; $h = 1/1 - t_o$, same as **ii**. above
**2.** $t_o + 2h = 1$; $h = 1/2 - ½\, t_o$, not binding
**3.** $t_o + 3h = 1$; $h = 1/3 - {}_{1/3}\, t_o$, not binding
**4.** $t_o + 4h = 1$; $h = 1/4 - ¼\, t_o$, not binding

**5.** $t_o + h = 2$; $h = 2/1 - t_o$
**6.** $t_o + 2h = 2$; $h = 2/2 - ½\, t_o$, not binding
**7.** $t_o + 3h = 2$; $h = 2/3 - {}_{1/3}\, t_o$, not binding
**8.** $t_o + 4h = 2$; $h = 2/4 - ¼\, t_o$, not binding

**9.** $t_o + h = 5$; $h = 5/1 - t_o$
**10.** $t_o + 2h = 5$; $h = 5/2 - ½\, t_o$
**11.** $t_o + 3h = 5$; $h = 5/3 - {}_{1/3}\, t_o$
**12.** $t_o + 4h = 5$; $h = 5/4 - ¼\, t_o$ same as **iii**. above

Lines (3b) lead to level set vertices in $\boldsymbol{D}_o$, Figure 2. ($D_o$ and seven shape level sets were drawn using Corel, plotting vertices. Shapes were calculated from the average of vertex ($t_o$, $h$) values for each SLS.) Computer programming and conceptualization motivates data object (4), a right ragged matrix, $\boldsymbol{S}$ rows and $(1 + \boldsymbol{K}_s + 1 + 2\boldsymbol{V}_s)$ entries in each row.

$$\{(\boldsymbol{K}_s, \nu_{s,k};\ \boldsymbol{V}_s, (t_o, h)_{s,v}) \mid s = 1 \text{ to } \boldsymbol{S},\ k = 1 \text{ to } \boldsymbol{K}_s,\ v = 1 \text{ to } \boldsymbol{V}_s\} \qquad (4)$$

$s$ indexes the shape level sets *and* shapes, $\boldsymbol{K}_s$ = the number of bins for the $s^{th}$ shape, $\nu_{s,k}$ = the $k^{th}$ bin count for the $s^{th}$ shape, $\boldsymbol{V}_s \equiv$ number of vertices for $s^{th}$ shape level set, $(t_o, h)_{s,v}$, is the $v^{th}$ vertex for the $s^{th}$ shape level set, $v = 1$ to $\boldsymbol{V}_s$, $s = 1$ to $\boldsymbol{S} \equiv$ number of shapes of at most a prescribed number of bins, $\boldsymbol{K}$. (That is, "$\nu_{s,k}$" with $s$, $k$ subscripts is a bin count for $s^{th}$ shape, $k^{th}$ bin. "$v$" without subscripts, as in "s,$v$" is a vertex index, $v = 1$ to $\boldsymbol{V}_s$, for the $s^{th}$ shape.) Table 2 shows object (4) ordered lexicographically first on $\boldsymbol{K}_s$, then on $\nu_{s,k}$.

**Table 2, Data object (4) for {1, 2, 5}**

$\{(\boldsymbol{K}_S, (\mathbf{v}_{s,k}); V_S, (t_o, h)_{S,V}), s = 1 \text{ to } 7, k = 1 \text{ to } \boldsymbol{K}_s, v = 1 \text{ to } V_s\}$

(**1**, (3); **3**, (1, 4), (1, 8), (-3, 8) )
(**2**, (1, 2); **4**, (-1, 3), (-6, 8), (-7, 8), (-3, 4) )
(**2**, (2, 1); **5**, (1, 2), (1, 4), (-3, 8), (-6, 8), (-1, 3) )
(**3**, (1, 1, 1); **4**, (.5, 1.5), (-1, 3), (-3, 4), (-1, 2) )
(**3**, (2, 0, 1); **4**, (1, 1.33), (1, 2), (-1, 3), (.5, 1.5) )
(**4**, (1, 1, 0, 1); **4**, (1, 1), (0.5, 1.5), (-1, 2), (-0.33, 1.33) )
(**4**, (2, 0, 0, 1); **3**, (1, 1), (1, 1.33), (.5, 1.5) )

These shape level sets are shown above as follows:

(**3**), Light Orange
3 ($t_o$,$h$) vertices: (1, **4**), (1, **8**), (-3, **8**), Avg($t_o$,$h$) = (-0.33, 6.67)
Min $h$ = **4**, Max $h$ = **8**

(**1, 2**), Purple
4 ($t_o$,$h$) vertices: (-1, **3**), (-6, **8**), (-7, **8**), (-3, **4**); Avg($t_o$,$h$) = (-4.25, 5.75)
Min $h$ = **3**, Max $h$ = **8**

(**2, 1**), Red
5 ($t_o$,$h$) vertices: (1, **2**), (1, **4**), (-3, **8**), (-6, **8**), (-1, **3**); Avg($t_o$,$h$) = (-1.6, 5)
Min $h$ = **2**, Max $h$ = **8**

(**1, 1, 1**), Black
4 ($t_o$,$h$) vertices: (.5, **1.5**), (-1, **3**), (-3, **4**), (-1, **2**); Avg($t_o$,$h$) = (-1.125,2.625)
Min $h$ = **1.5**, Max $h$ = **4**

(**2, 0, 1**), Green
4 ($t_o$,$h$) vertices: (1, **1.33**), (1, **2**), (-1, **3**), (.5, **1.5**); Avg($t_o$,$h$) = (0.375,1.958)
Min $h$ = **1.33**, Max $h$ = **3**

(**1, 1, 0, 1**), Dark Orange
4 ($t_o$,$h$) vertices: (1, **1**), (0.5, **1.5**), (-1, **2**), (-0.33, **1.33**); Avg($t_o$,$h$) = (0.042, 1.458)
Min $h$ = **1**, Max $h$ = **2**

(**2, 0, 0, 1**), Light Blue
3 ($t_o$,$h$) vertices: (1, **1**), (1, **1.33**), (.5, **1.5**); Avg($t_o$,$h$) = (0.833, 1.278)
Min $h$ = **1**, Max $h$ = **1.5**

## 2.4 Mean and Variance Consistency of Shapes

We need to step back from constraints (1), (2ab) and look at shape level sets. Table 1 is arises from easy determination of *joint* consistency. Shapes are MV jointly consistent, or not. However, *not jointly* MV *consistent* happens in four ways: *Individually-but-not-jointly mean and variance consistent*; *Mean-consistent-but-not-variance consistent*; *Variance-consistent-but-not-mean consistent*; *Neither mean nor variance consistent*. For each shape we should know if it is jointly mean and variance consistent, and if not, individual consistencies.

Straight lines (1), (2a) can be graphed in $\boldsymbol{D}_o$. *Individual* mean or variance consistency of a shape is indicated by intersection of line (1) or line (2a) with the associated shape level set. If a mean (or variance) constraint line intersects its shape level set, then the shape is individually mean (or variance) consistent with the sample mean (or variance).

For *not* jointly mean and variance consistent shapes, *individual* mean and variance consistency is determined separately from sign changes at level set vertices for a mean constraint function, $f_m(t_o, h) \equiv \bar{x}_g - \bar{x}_x$, and a variance constraint function, $f_v(t_o, h) \equiv s^2_g - s^2_x$. A sign change associated with level set vertices indicates $f(t_o, h) = 0$ inside a level set indicating mean or variance consistency of a shape. Individual consistency of both mean and variance consistency together does *not* imply joint consistency. Visualizing straight-line graphs of constraints (1), (2a) on Fig. 2 can illustrate these possibilities. Although variance and mean constraints necessarily intersect, they may or may not intersect their shape level sets, and may or may not intersect *inside* their shape level sets.

That is,

- If (1), (2a) intersect ***inside*** the SLS for $v_k$, then the intersection $t_o$, $h$ values lead to the same $v_k$ bin counts used in (1), (2a). Then the *shape* $v_k$ is MV *jointly consistent*. (Otherwise, (1), (2a) intersect each other outside of a shape level set, regardless of their intersections with the shape level set.)

- Neither (1) nor (2a) necessarily intersect the SLS for $v_k$. Each may or may not intersect a shape level set, associated with individual mean, variance consistency or inconsistency.

There are five possibilities altogether, summarized below.

## Notation for MV Shape Consistency

$\mathbf{J_g}$ ≡ Jointly mean and variance consistent shapes, situation 1, below.
$\mathbf{M_g}$ ≡ Mean consistent shapes, situations 1, 2, 3; not 4, 5.
$\mathbf{V_g}$ ≡ Variance consistent shapes, situations 1, 2, 4; not 3, 5.
**S** ≡ All 123 shapes of at most six uniform width bins for $x_i$, situations 1–5.

1. $\mathbf{J_g}$: As already noted, MV constraint lines, (1), (2a), may intersect each other *inside* the SLS for $v_k$. Then $v_k$ is <u>*jointly* MV consistent</u>. Table 1 shows three examples: (**3,4,5**), (**5,3,4**), (**1,2,3,1,2,3**). Looking ahead, Table 3 shows **8 jointly consistent shapes**, denoted "$\mathbf{J_g}$".
2. $(\mathbf{M_g} \cap \mathbf{V_g})/\mathbf{J_g}$: Lines (1), (2a) may intersect the $v_k$ SLS, but *intersect each other outside* of this SLS. Then $v_k$ is <u>***individually** but not jointly* MV consistent</u>. Table 1 shows one such example, (**5,4,3**). Table 3 shows **11** *individually* but not jointly MV consistent shapes.

3. $\mathbf{M_g/V_g}$: Line (2a) but not line (1) may intersect the SLS. Then $v_k$ is mean consistent but not variance consistent, Table 3 **$\mathbf{M_g/V_g}$ - 10 shapes**, for example, (**1,8,3**), (**1,10,1**) but not (**2,5,5**).
4. $\mathbf{V_g/M_g}$: Line (1) but not line (2a) may intersect the SLS. Then $v_k$ is variance consistent but not mean consistent, Table 3, **$\mathbf{V_g/M_g}$ - 2 shapes**, for example (**4,3,5**), (**4,4,4**), but not (**4,5,3**).
5. **S/($\mathbf{M_g \cup V_g}$)**: If neither line (1) nor line (2a) intersects the level set, then $v_k$ is neither variance nor mean consistent. Table 3 shows at most three bins. There are 123 shapes of at most six bins, so **92 shapes** *not* shown in Table 3. In Table 1, shapes (**1,2,3,3,2,1**), (**3,2,1,1,2,3**) are neither mean nor variance consistent. Table 3*, Appendix D, shows situations 1-4 for at most six bins.

The five situations above partition uniform bin width histogram shapes, **S**.

Figure 2 shows seven shape level sets identifying seven shapes of at most four bins for {1, 2, 5}. Sample mean and variance are $\bar{x}_x = 8/3$, $s^2_x = 13/3$ leading to seven variance constraints, (1) and seven mean constraints (2b). Visualizing intersections of mean and variance constraint lines with the shape level sets illustrates situations 1 – 5 above for each shape. Of course, actual consistency is not determined graphically. Table 1 and discussion shows that joint mean and variance consistency is easily determined. Again, not jointly consistent shapes are tested separately for mean consistency, $\mathbf{M_g}$, and variance consistency, $\mathbf{V_g}$, via sign changes in $\bar{x}_g - \bar{x}_x$, and $s^2_g - s^2_x$ at level set vertices.

Also, recall that for each shape, the MISE and maximum likelihood histograms densities occur for minimum bin width for a shape. This is easily seen at vertices with minimum value on the vertical bin width axis. Further, by projecting shape level set bin width minima and maxima to the vertical bin width axis, we obtain the partition of bin width values into cells associated with a fixed set of shapes. This exactly implements histogram bin width shape stability criterion articulated by Simonff & Udina (1997).

Construction of shape level sets may be unfamiliar or appear complicated until implemented in software. Calculating shape level sets is robust. A simple feature is that shape level set vertices are determined with a single calculation, not approximated iteratively. There is iteration or looping through the boundary lines (3b), but each vertex involves a single calculation for $t_o$ and $h$. We have explored distinctly inferior computational procedures: 1. an awkward adaptation of the linear programing simplex algorithm; 2. grid search; 3. non-linear solver suites, such as "What's Best?," LINDO systems.

Data {1, 2, 5} clarifies shape level sets. Returning to Table 1 data, Table 3 shows variance and mean *consistency*, as well as grouped data skewness.

**2.5 An Example**. Table 3 breaks out mean or variance consistent shapes of at most three bins, $\mathbf{M_g} \cup \mathbf{V_g}$ (for Table 1 data) into columns: $\mathbf{M_g}$, $\mathbf{V_g}$, $\mathbf{M_g} \cap \mathbf{V_g}$, $\mathbf{J_g}$. Shapes (**1,2,3,3,2,1**) and (**3,2,1,1,2,3**), Table 1, are situation 5, in $\mathbf{S}/(\mathbf{M_g} \cup \mathbf{V_g})$, are neither mean nor variance consistent and *not* in Table 3. In Table 3, shapes are listed lexicographically on $\boldsymbol{K}$ (not shown), then bin counts. ("MISE" in Table 3 abbreviates *mean integrated squared error*. Rudemo, 1982; Scott, 1992; "ML" *maximum likelihood*).

**Table 3** $M_g \cup V_g \supset$ $M_g \cap V_g$ $\supset \neq J_g$ $T_{FP} \supset$ $F_{FP}$ $T_{FP} \cap J_g$

31 Mean or Variance consistent shapes, out of 123*

| $M_g$: Mean Consistent Shapes | $V_g$: Variance Consistent Shapes | $M_g \cap V_g$ Shapes | $J_g$: Mean & Variance **Jointly Consistent** | **Fisher-Pearson Skewness -0.0288** **Shape Skewness** | **Skewness Rank of Shapes Within*** **$T_{FP}$: Ten%** of $g_x$ | **Skewness Rank of Shapes Within**** **$F_{FP}$: Five%** of $g_x$ | $T_{FP} \cap J_g$ |
|---|---|---|---|---|---|---|---|
| 12 | 12 | 12 | 12 exact MISE | 0 | | | |
| 1,11 | 1,11 | 1,11 | 1,11 | | | | |
| 2,10 | 2,10 | 2,10 | | | | | |
| 3,9 | 3,9 | 3,9 | 3,9 | | | | |
| 4,8 | 4,8 | 4,8 | | | | | |
| 5,7 | 5,7 | 5,7 | | | | | |
| 6,6 | 6,6 | 6,6 | 6,6 Rice,Sh MISE | 0 | 5 | 5 | 6,6 |
| 7,5 | 7,5 | 7,5 | 7,5 | | | | |
| 8,4 | | | | | | | |
| 9,3 | | | | | | | |
| 10,2 | | | | | | | |
| 11,1 | 11,1 | 11,1 | | | | | |
| 1,6,5 | 1,6,5 | 1,6,5 | 1,6,5 | | | | |
| 1,8,3 | | | | | | | |
| 1,10,1 | | | | | | | |
| 2,5,5 | 2,5,5 | 2,5,5 | | | | | |
| 2,7,3 | | | | -0.075 | -8 | | |
| 3,4,5 | 3,4,5 | 3,4,5 | 3,4,5 ML | | | | |
| 3,5,4 | 3,5,4 | 3,5,4 | | | | | |
| 3,6,3 | 3,6,3 | 3,6,3 | | 0 | 5 | 5 | |
| 3,7,2 | | | | | | | |
| 3,8,1 | | | | -0.0548 | -4 | -4 | |
| | 4,3,5 | | | | | | |
| | 4,4,4 | | | 0 | 5 | 5 | |
| 4,5,3 | 4,5,3 | 4,5,3 | | | | | |
| 5,3,4 | 5,3,4 | 5,3,4 | 5,3,4 | | | | |
| 5,4,3 | 5,4,3 | 5,4,3 | | | | | |
| 5,5,2 | | | | | | | |
| 6,3,3 | 6,3,3 | 6,3,3 | | | | | |
| 6,4,2 | | | | | | | |
| 6,5,1 | 6,5,1 | 6,5,1 | | | | | |

**3. Fisher-Pearson skewness, Fisher-Pearson adjusted skewness; Mode inversion**

Fisher-Pearson population skewness (*FPS*; $FPS_{\mathbf{x}}$, $FPS_{\mathbf{g}}$), (5ab), and Fisher-Pearson adjusted sample skewness (*FPAS*; $FPAS_{\mathbf{x}}$, $FPAS_{\mathbf{g}}$), (5cd), are translation and scale invariant standardized third moments (various, including Doane, Seward, 2011; Groeneveld, Meeden, 1984). Unlike variance and mean constraints, skewness constraints (5abcd) depend on $t_o$, $h$ only through bin counts, $v_k$. Also, bin counts and histogram density function values differ by the factors $nh$, $1/nh$, so their skewness constraints are identical.

$$\textit{Fisher-Pearson skewness} \equiv FPS_x \equiv \frac{\frac{1}{n}\sum_{i=1}^{n}(x_i - \bar{x})^3}{(\frac{1}{n}\sum_{i=1}^{n}(x_i - \bar{x})^2)^{3/2}} \quad (5.a)$$

$$\textit{Grouped data Fisher-Pearson skewness} \equiv FPS_g\ (v_k, \boldsymbol{K}, n) \equiv \frac{\frac{1}{n}\sum_{k=1}^{K} v_k (k - \bar{k})^3}{(\frac{1}{n}\sum_{k=1}^{K} v_k (k - \bar{k})^2)^{3/2}} \sim= FPS \quad (5.b)$$

$$\textit{Fisher-Pearson adjusted skewness} \equiv FPAS_x \equiv \frac{n}{(n-1)(n-2)}\sum_{i=1}^{n}[(x_i - \bar{x})/s]^3 \quad (5.c)$$

$$\textit{Grouped data Fisher-Pearson adjusted skewness} \equiv FPAS_{\mathbf{g}} = \frac{n^{3/2}(n-1)^{1/2}}{(n-1)(n-2)} \bullet g_{\mathbf{g}} \sim= FPAS_{\mathbf{x}} \quad (5.d)$$

$$FPAS_{\mathbf{x}} = \frac{n^{3/2}(n-1)^{1/2}}{(n-1)(n-2)} \bullet FPS_{\mathbf{x}} > FPS_{\mathbf{x}} \quad (5.e)$$

Since these grouped data measures of skewness depend only on shape, equality with data skewness is rare, unlike mean and variance consistency. However, we can rank shapes via deviation objective functions such as
$f_{FPS}\ (t_o, h; x_i) \equiv |\ FPS_{\mathbf{g}}(t_o, h; x_i) - FPS_{\mathbf{x}}(x_i)\ |, f_{FPAS}\ (t_o, h; x_i) \equiv |\ FPAS_{\mathbf{g}}(t_o, h; x_i) - FPAS_{\mathbf{x}}(x_i)$.
Monotone relationship, (5.e), connects Fisher-Pearson adjusted skewness, $FPAS_x$ to Fisher-Pearson skewness, $FPS_x$, and implies that rankings according to closeness of shape skewness, $FPS_g$, to data skewness, $FPS_x$; or adjusted $FPAS_g$ to $FPAS_x$, are the same. Consequently, the same *skewness-good* histogram shapes emerge from all four combinations of *FPS*, *FPAS* and frequency histograms, density histograms. (That is, simplifying *FPS* and *FPSA* expressions with bin frequencies, $v_k$, and bin midpoints, $t_o + (k - ½)h$, or integrating a density, $v_k/nh$, eliminates occurrences of $t_o$, $h$ outside of $v_k(t_o, h; x_i)$. Consequently, on SLSs, standardized third moment *FPS* and *FPAS* skewness are constant so shape level sets also are *FPS* and *FPAS* level sets.

Since histogram skewness *rarely* equals data skewness, histogram shapes are ranked relative to data skewness by FPS or FPAS histogram skewness. Table 3 shows columns $\mathbf{T_{FP}}$, $\mathbf{F_{FP}}$, for, respectively, the 10%, 5% *of all* 123 *shapes* greater than *and* 10%, 5% less than data Fisher-Pearson skewness. Skewness rankings include shapes that are neither mean nor variance consistent. Table 3, columns $\mathbf{T_{FP}}$, $\mathbf{F_{FP}}$ show only three bin shapes for situations 1-4, $\mathbf{M_g} \cup \mathbf{V_g}$. $\mathbf{T_{FP} \cap J_g}$ shows $\mathbf{T_{FP}}$ shapes that are MV jointly consistent and among the 10% of shapes closer to data FP skewness.

That is, Data #3, Table 1, Weber (2008a), has 123 shapes of at most six uniform width bins. Table 3 shows only shapes of at most three bins, and a similar table in an appendix shows shapes with six bins. $\mathbf{T_{FP}}$ has shapes ranked 1 to 12 above data FP skewness and –1 to –12 below; $\mathbf{F_{FP}}$, 1 to 6, –1 to –6. Large intervals of ranks –12 to +12, –6 to + 6 were needed for Table 3 because many shapes that are close in skewness to the data are neither mean nor variance consistent. Finally, $\mathbf{T_{FP} \cap J_g}$ indicates jointly consistent shapes with FP skewness rank within –12 to +12 relative to the data FP skewness. Many shapes that are FP skewness ranked –12 to +12 are neither mean nor variance consistent, not shown in Table 3.

We do not explore mode inversion beyond showing that this can occur, Table 4.

| **Table 4** | Bin counts ("Shape") | $t_0$ | $h$ | $x_i$ |
|---|---|---|---|---|
| Table 1 | **1, 2, 3, 3, 2, 1** | 0.0000 | 1.2500 | .37, 1.13, 1.23, 2.25, 2.35, 2.45, 3.37, 4.37, |
| *n* =12 | **3, 2, 1, 1, 2, 3** | 1.3600 | 1.0000 | 4.47, 5.37, 5.47, 5.61 |
| Appendix A | **1, 9, 9, 1** | -0.680 | 2.8400 | 2.05, 2.27, 2.50, 2.95, 3.18, 3.41, 3.64, 3.86, 4.09, 4.32 |
| *n* =20 | **6, 4, 4, 6** | 1.9542 | 1.5229 | 5.68, 5.91, 6.14, 6.36, 6.59, 6.82, 7.05, 7.50, 7.73, 7.95 |

Uni-modal and bi-modal inversion means histogram variance, kurtosis and modes can be misleading if histograms are not compared with sample variance, kurtosis. (Mode inversion can change skewness, however Table 4 shapes are symmetric, skewness is the same. Table 1 data were constructed to have mode inversions (1,2,3,3,2,1), (3,2,1,1,2,3). Shapes (1,9,9,1), (6,4,4,6) are symmetric for exactly symmetric Appendix A data.

## 4. Summary, Conclusion, other remarks.

Defining good histograms by varying the bins, focusing on numbers of bins or bin width dates at least from Sturges, (1926). This contrasts with making the best use of pre-binned data, circa 1900, Sheppard, (1898), K. Pearson (various), Smith, 1916, Fisher, (1916), Correspondence, Pearson, E. (1968); Stigler, (2005), (also see Hollerith, circa 1890); and recently Minnotte, (1998). Among many efforts since 1950, MISE and Simonoff & Udina (1997) bin width shape stability come to mind.

In contrast, MOM corresponds to the way smaller sample histograms often are used as a data graphic showing skewness, variance and central tendency. We do not believe that smaller samples should be used for histogram density estimators because step function densities are rarely used. Apparently, this was K. Pearson's view (regardless of sample size), since his parametric frequency curves do not include histogram densities.

Even though the elementary mathematics of shape level sets is detailed and tedious (like elementary linear programming), MOM *concepts* (agreement of data and histogram mean, variance, and skewness) are compelling to introductory statistics students, many statisticians and data scientists. MOM is easily explained to anyone already familiar with mean, variance and skewness.

Weber (2016, 2008a) present relevant discussions. Weber (2008a) suggests that skewness can be satisfied exactly without distinguishing among non-central, central, and standardized third moment approaches to skewness. This is partly true, however, as noted, this is not true for Fisher-Pearson skewness. Since Fisher-Pearson skewness rarely is satisfied exactly, why satisfy mean and variance exactly? Instead minimize a composite deviation of all three, using mean, variance, and skewness in a conceptually familiar analogue of least squares, or pursue less familiar non-central or central (but not standardized) moments.

Hoaglin, Mosteller, Tukey (1983, 2000); McNeil (1977) (pp 3-6) suggest that stem-and-leaf plots are superior to histograms. Stem-and-leaf displays may be easier to draw on a blackboard, however that was then, this is now. Further, McNeil (1977) observes that vertical bin locations and uniform widths need not always be a power of ten and further points out that the shapes of stem-and-leaf plots are highly variable. With these generalizations, *stem-and-leaf plots are simply histograms rotated 90 degrees*.

Everything we have stated about histograms applies to stem-and-leaf plots. The argument for MOM histograms applies equally for MOM stem-and-leaf plots. With an example (pp 9-11) McNeil also points out that transformations of the data together with stem-and-leaf plots often can identify structures that are difficult to perceive without transforming the data. Clearly the same detective work can be done with histograms. MISE, maximum likelihood, and Simonoff & Udina all require minimum and maximum bin width (often infima and suprema) for shapes and these are easily available from shape level set vertices in $\{(t_o,h)\}$. Apparently other procedures do not exactly determine these extreme values.

**Acknowledgements.** To be provided at the time of publication.

## Appendix A. Additional details on Shape Skewness Reversal

Shape reversals matter on account of two unexpected features: **1**. Some histogram shapes for symmetric data are asymmetric; **2**. For translated bins of the same width, asymmetric shapes for symmetric data have a reversal shape and opposite skewness.

This is emphasized by Table A.

**Table A: Histogram Shape Reversals, A-H**

| Example Data Set # | Shape | Bin Location $t_0$ | Bin Width $h$ | Symmetric data |
|---|---|---|---|---|
| **2** | **A: 10, 9, 1** | 1.4250 | **3.2075** | 2.05, 2.27, 2.50, 2.95, 3.18, 3.41, 3.64, 3.86, 4.09, 4.32 |
| *n*=20 | **B: 1, 9, 10** | -1.048 | **3.2075** | 5.68, 5.91, 6.14, 6.36, 6.59, 6.82, 7.05, 7.50, 7.73, 7.95 |
| | **C: 8, 4, 7, 1** | 1.9767 | **1.9789** | |
| | **D: 1, 7, 4, 8** | 0.1078 | **1.9789** | |
| | **E: 6, 4, 4, 5, 1** | 1.9829 | **1.4750** | |
| | **F: 1, 5, 4, 4, 6** | 0.6421 | **1.4750** | |
| | **G: 4, 6, 0, 5, 4,1** | 1.9619 | **1.9060** | |
| | **H: 1, 4, 5, 0, 6, 4** | 0.8944 | **1.9060** | |

We may be misled by unlucky choice of histogram bins that show skewness opposite of data skewness; or asymmetry for data that is approximately symmetric. How widely known is this? Which, if any, statistical software suites block or flag histograms with skewness sign opposite calculated data skewness? R. J. Little (2013) reports that histograms are the most mentioned significant simple statistical idea or tool, mentioned 12 times by 30 respondents, compared to 7 for the second most mentioned statistical idea or tool. So histograms are very much in the minds of statisticians. Histograms A-H above are not symmetric and rule out the possibility that symmetric data have only symmetric histograms, as is occasionally suggested.

Reversal of shape simply from data symmetry and bin translation is clarified by the following lemma.

**Data Symmetry and Shape Reversal Lemma**.

**A:** Data are exactly symmetric iff

**B:** For every uniform bin width histogram shape, the reversal shape occurs with different uniform width bins, including among others, translations of the bins.

**Proof**:

■ **A** $\Rightarrow$ **B** Consider exactly symmetric data points, a set of *uniform width* bins, edges adjusted to not equal data values, and histogram shape. Reflecting data points and bin edges *across the data mean leads to the reversal shape*, for the same data values and value frequencies, since data points are exactly symmetric about the data mean. (Since bins are uniform width, the reflected bin edges can be obtained via a translation.) ■

■ **B** $\Rightarrow$ **A** Conversely, suppose that every shape of a uniform bin width histogram is accompanied by its reversal, for different uniform width bins. Isolate data values with small bin width. Since the reversal shape also occurs, the data *value **frequencies** are exactly symmetric*. Further, as the bin width becomes arbitrarily small, absolute differences between the data mean and values in bins located symmetrically above and below the bin containing the data mean is of the order of the bin width and becomes arbitrarily small. Thus, the *data **values** are exactly symmetric* about the data mean.

*Symmetric value **frequencies** and symmetric **values** is exact data symmetry*. ■

This proves the **Data Symmetry and Shape Reversal Lemma**. Examples A–H, Table A, show that exactly symmetric data can have asymmetric uniform bin width histogram shapes.

**Appendix B. Exact implementation of Simonoff-Udina (1997) Bin Width Shape Stability Criterion**

Shape level set vertices provide minimum and maximum bin widths for shapes, $h_{min}$ and $h_{max}$, leading to a succinct (elegant?) exact rendering of Simonoff & Udina (1997) shape stability criteria. Although focusing on bin width with the minimum shape variability due to translation seems attractive, we do not think that by itself that insures "good" histograms any more than other bin width rules. Nevertheless, shape level sets lead to exact description of shape variability dependence on bin width for uniform bin width histograms of at most four bins for {1, 2, 5}, Figure 2.

The set {min-*h*, max-*h* | for shapes} → {1, 2**;** 1, 1.5**;** 4, 8**;** 3, 8**;** 2, 8**;** 1.5, 4**;** 1.33, 3} →

{**1, 1.33, 1.5, 2, 3, 4, 8**}

By inspection, shape stability cells are:
(1, 1.33) – **two shapes: (2, 0, 0, 1), (1, 1, 0, 1)**
(1.33, 1.5) – **three shapes: (2, 0, 0, 1), (1, 1, 0, 1), (2, 0, 1)**
(1.5, 2) – **three shapes: (1, 1, 0, 1), (2, 0, 1), (1, 1, 1)**
(2, 3) – **three shapes: (2, 0, 1), (1, 1, 1), (2, 1)**
(3, 4) – **three shapes: (1, 1, 1), (2, 1), (1, 2)**
(4, 8) – **three shapes: (2, 1), (1, 2), (3)**

So best bin widths according to Simonoff, Udina shape stability are values in the interval (1, 1.33).

Shape level sets involved here and Figure 2, for {1, 2, 5} for at most four bins are as follows:

(**3**), Light Orange
3 ($t_o$,$h$) vertices: (1, **4**), (1, **8**), (-3, **8**), Avg($t_o$,$h$) = (-0.33, 6.67)
Min $h$ = **4**, Max $h$ = **8**

(**1, 2**), Purple
4 ($t_o$,$h$) vertices: (-1, **3**), (-6, **8**), (-7, **8**), (-3, **4**); Avg($t_o$,$h$) = (-4.25, 5.75)
Min $h$ = **3**, Max $h$ = **8**

(**2, 1**), Red
5 ($t_o$,$h$) vertices: (1, **2**), (1, **4**), (-3, **8**), (-6, **8**), (-1, **3**); Avg($t_o$,$h$) = (-1.6, 5)
Min $h$ = **2**, Max $h$ = **8**

(**1, 1, 1**), Black
4 ($t_o$,$h$) vertices: (.5, **1.5**), (-1, **3**), (-3, **4**), (-1, **2**); Avg($t_o$,$h$) = (-1.125,2.625)
Min $h$ = **1.5**, Max $h$ = **4**

(**2, 0, 1**), Green
4 ($t_o$,$h$) vertices: (1, **1.33**), (1, **2**), (-1, **3**), (.5, **1.5**); Avg($t_o$,$h$) = (0.375,1.958)
Min $h$ = **1.33**, Max $h$ = **3**

(**1, 1, 0, 1**), Dark Orange
4 ($t_o$,$h$) vertices: (1, **1**), (0.5, **1.5**), (-1, **2**), (-0.33, **1.33**); Avg($t_o$,$h$) = (0.042, 1.458)
Min $h$ = **1**, Max $h$ = **2**

(**2, 0, 0, 1**), Light Blue
3 ($t_o$,$h$) vertices: (1, **1**), (1, **1.33**), (.5, **1.5**); Avg($t_o$,$h$) = (0.833, 1.278)
Min $h$ = **1**, Max $h$ = **1.5**

Also, Appendix C explains matching all grouped data moments, including the following: for small $h = h_m = 1/mQ$, $m = 1$ to $\infty$, $P_{m,i}^* = m P_i$ and $x_i = P_{m,i}^* (1/mQ) = P_{m,i}^* h_m$, there will be only one shape regardless of location. Translations all lead to the same shape. So, these point values for $h$ are the most shape stable bin widths, although these histograms are simply dot plots. Like MISE and maximum likelihood, a rationale for upper and lower bounds on bin width is needed. MISE, maximum likelihood and bin width shape stability are not enough without bin width constraints.

## Appendix C. Matching ALL moments

For uniform width bins for which mid points equal data values the frequency histogram grouped data moments are identical to data moments.

More explicitly, real world data, $x_i$, are represented by rational numbers, $x_i = p_i/q_i$ where $p_i$, $q_i$ are relatively prime integers. Define $Q$ = the least common multiple of the integers $q_i$, i.e. $Q \equiv LCM\{q_i \mid i = 1 \text{ to } n\}$, so $k_i q_i = Q$, and $x_i = p_i/q_i = k_i p_i/k_i q_i = P_i/Q$.

Consider bin widths $h_m = 1/mQ$, $m = 1$ to $\infty$, $P_{m,i}{}^* = m\, P_i$ and $x_i = P_{m,i}{}^* \,(1/mQ) = P_{m,i}{}^*\, h_m$.

Define $t_o = (x_{\min} - \frac{1}{2}\, h_m)$. This gives bins so that every data value is the midpoint of the bin that contains it. So, frequency histogram *grouped* data moments are the same as data moments, for $m = 1$ to $\infty$. Of course, at this point, "grouped data" is a misnomer since small bin widths that isolate data values, do not *group* data values. A result emailed from Scott, circa 2012, can be revisited. Scott: "…As $h \rightarrow$ zero, all moments converge to data moments. …" Our comment: $h$ does not need to go to zero to achieve *exact* agreement of all grouped data *frequency* histogram moments with data moments. Histogram *densities* add $h^2/12$ to variance. Obviously $h^2/12 \rightarrow$ zero as $h \rightarrow$ zero however histogram densities become unbounded as $h \rightarrow$ zero.

## Appendix D. Table 3*, augmented Table 3 showing six or fewer bins

**Table 3*-i** $M_g \cup V_g \supset$ $M_g \cap V_g$ $\supset \neq J_g$ $T_{FP} \supset$ $F_{FP}$ $T_{FP} \cap J_g$

79 Mean or Variance consistent shapes, out of 123.

| **$M_g$**: Mean Consistent Shapes | **$V_g$**: Variance Consistent Shapes | **$M_g \cap V_g$** Shapes | **$J_g$: Mean & Variance Jointly Consistent** | **Fisher-Pearson Data Skewness -0.0288** Shape Skewness | **Skewness Rank of Shapes Within*** $T_{FP}$: Ten% of $g_x$ | **Skewness Rank of Shapes Within**** $F_{FP}$: Five% of $g_x$ | $T_{FP} \cap J_g$ |
|---|---|---|---|---|---|---|---|
| 12 | 12 | 12 | 12 **exact MISE** | **0** | | | |
| 1,11 | 1,11 | 1,11 | 1,11 | | | | |
| 2,10 | 2,10 | 2,10 | | | | | |
| 3,9 | 3,9 | 3,9 | 3,9 | | | | |
| 4,8 | 4,8 | 4,8 | | | | | |
| 5,7 | 5,7 | 5,7 | | | | | |
| 6,6 | 6,6 | 6,6 | 6,6 **Rice,Sh MISE** | 0 | **5** | **5** | **6,6** |
| 7,5 | 7,5 | 7,5 | 7,5 | | | | |
| **8,4** | | | | | | | |
| 9,3 | | | | | | | |
| 10,2 | | | | | | | |
| 11,1 | 11,1 | 11,1 | | | | | |
| 1,6,5 | 1,6,5 | 1,6,5 | 1,6,5 | | | | |
| **1,8,3** | | | | | | | |
| 1,10,1 | | | | | | | |
| 2,5,5 | 2,5,5 | 2,5,5 | | | | | |
| 2,7,3 | | | | -0.075 | **-8** | | |
| **3,4,5** | **3,4,5** | **3,4,5** | **3,4,5 ML** | | | | |
| 3,5,4 | 3,5,4 | 3,5,4 | | | | | |
| 3,6,3 | 3,6,3 | 3,6,3 | | 0 | **5** | **5** | |
| 3,7,2 | | | | | | | |
| **3,8,1**[5] | | | | **-0.0548** | **-4** | **-4** | |
| | **4,3,5** | | | | | | |
| | 4,4,4 | | | 0 | **5** | **5** | |
| 4,5,3 | 4,5,3 | 4,5,3 | | | | | |
| **5,3,4** | **5,3,4** | **5,3,4** | **5,3,4** | | | | |
| **5,4,3** | **5,4,3** | **5,4,3** | | | | | |
| 5,5,2 | | | | | | | |
| 6,3,3 | 6,3,3 | 6,3,3 | | | | | |
| 6,4,2 | | | | | | | |
| 6,5,1 | 6,5,1 | 6,5,1 | | | | | |
| 1,2,4,5 | 1,2,4,5 | 1,2,4,5 | | | | | |
| 1,3,3,5 | 1,3,3,5 | 1,3,3,5 | 1,3,3,5 | | | | |
| 1,4,2,5 | 1,4,2,5 | 1,4,2,5 | | | | | |
| 1,5,1,5 | 1,5,1,5 | 1,5,1,5 | **1,5,1,5** | -0.0762 | **-9** | | **1,5,1,5** |
| 1,5,2,4 | 1,5,2,4 | 1,5,2,4 | | | | | |
| 1,5,3,3 | 1,5,3,3 | 1,5,3,3 | 1,5,3,3 | | | | |
| 1,5,4,2 | | | | | | | |
| 1,5,5,1 | 1,5,5,1 | 1,5,5,1 | **1,5,5,1** | 0 | **5** | **5** | **1,5,5,1** |
| | 2,4,1,5 | | | | | | |
| | 3,4,2,4 | | | | | | |
| 2,4,3,3 | 2,4,3,3 | 2,4,3,3 | | | Almst in $T_{FP}$ | | |

**Table 3*-ii** $M_g \cup V_g \supset$ $M_g \cap V_g$ $\supset \neq J_g$ $T_{FP} \supset$ $F_{FP}$ $T_{FP} \cap J_g$

79 Mean or Variance consistent shapes, out of 123.

| $M_g$: Mean Consistent Shapes | $V_g$: Variance Consistent Shapes | $M_g \cap V_g$ Shapes | $J_g$: Mean & Variance **Jointly Consistent** | **Fisher-Pearson Data Skewness -0.0288** <br> **Shape Skewness** | **Skewness Rank of Shapes Within*** <br> **$T_{FP}$: Ten%** of $g_x$ | **Skewness Rank of Shapes Within*** <br> **$F_{FP}$: Five%** of $g_x$ | $T_{FP} \cap J_g$ |
|---|---|---|---|---|---|---|---|
| 3,3,1,5 | 3,3,1,5 | 3,3,1,5 | | | | | |
| 3,3,2,4 | 3,3,2,4 | 3,3,2,4 | | -0.0491 | **-3** | **-3** | |
| 3,3,3,3 | 3,3,3,3 | 3,3,3,3 | **3,3,3,3** | 0 | **5** | **5** | **3,3,3,3** |
| 3,4,2,3 | 3,4,2,3 | 3,4,2,3 | | | | | |
| 3,4,3,2 | 3,4,3,2 | 3,4,3,2 | | | | | |
| 3,4,4,1 | 3,4,4,1 | 3,4,4,1 | 3,4,4,1 | | | | |
| 1,2,3,1,5 | 1,2,3,1,5 | 1,2,3,1,5 | 1,2,3,1,5 | | | | |
| 1,2,3,2,4 | 1,2,3,2,4 | 1,2,3,2,4 | | | | | |
| 1,2,3,3,3 | 1,2,3,3,3 | 1,2,3,3,3 | 1,2,3,3,3 | | | | |
| 1,2,4,2,3 | 1,2,4,2,3 | 1,2,4,2,3 | 1,2,4,2,3 | | | | |
| 1,2,4,4,1 | 1,2,4,4,1 | 1,2,4,4,1 | | | | | |
| 1,3,3,2,3 | 1,3,3,2,3 | 1,3,3,2,3 | | | | | |
| 1,3,3,4,1 | 1,3,3,4,1 | 1,3,3,4,1 | | | | | |
| 1,4,2,2,3 | 1,4,2,2,3 | 1,4,2,2,3 | | | | | |
| 1,4,2,4,1 | 1,4,2,4,1 | 1,4,2,4,1 | **1,4,2,4,1** | 0 | **5** | **5** | **1,4,2,4,1** |
| 1,5,1,3,2 | | | | | | | |
| 1,5,1,4,1 | 1,5,1,4,1 | 1,5,1,4,1 | 1,5,1,4,1 | | | | |
| 2,3,2,2,3 | 2,3,2,2,3 | 2,3,2,2,3 | | | | | |
| 2,4,1,2,3 | | | | | | | |
| 2,4,1,3,2 | | | | | | | |
| | 3,2,2,2,3 | | | 0 | **5** | **5** | |
| 3,3,1,2,3 | 3,3,1,2,3 | 3,3,1,2,3 | | | | | |
| 3,3,1,3,2 | | | | | | | |
| 3,3,1,4,1 | 3,3,1,4,1 | 3,3,1,4,1 | | | | | |
| 1,2,3,1,2,3 | 1,2,3,1,2,3 | 1,2,3,1,2,3 | **1,2,3,1,2,3** | -0.0552 | **-6** | | **1,2,3,1,2,3** |
| 1,2,3,1,3,2 | 1,2,3,1,3,2 | 1,2,3,1,3,2 | | -0.0859 | **-11** | | |
| 1,2,3,1,4,1 | 1,2,3,1,4,1 | 1,2,3,1,4,1 | | | | | |
| | 1,2,3,2,3,1 | | | | | | |
| 2,1,3,1,2,3 | 2,1,3,1,2,3 | 2,1,3,1,2,3 | | | | | |
| | 2,1,3,2,2,2 | | | | | | |
| | 2,2,2,2,3,1 | | | | | | |
| 2,2,3,1,3,1 | 2,2,3,1,3,1 | 2,2,3,1,3,1 | | | | | |
| 3,0,3,1,2,3 | 3,0,3,1,2,3 | 3,0,3,1,2,3 | **( Exact ML )** | | | | |
| 3,1,3,1,2,2 | | | | | | | |
| | 3,1,3,2,2,1 | | | | | | |
| | 3,2,2,2,2,1 | | | | | | |
| | 3,3,1,2,2,1 | | | | | | |

* $T_{FP}$ means +/- 10% of 123 shapes.
That is, twelve shapes less than and twelve greater than the data Fisher-Pearson skewness

* $F_{FP}$ means +/- 5% of 123 shapes
That is, six shapes less than and six greater than the data Fisher-Pearson skewness

**Appendix E. List of partitions.**

Although "partitions" are a familiar analytic tool, it be helpful to list all that are used, mentioned here, or of possible interest:

1. Partition of a domain $\boldsymbol{D}_o$ of the space $(t_o,h)$ into shape level sets. ($\boldsymbol{D}_o$ defined so that the first bin, $[t_o + (k–1)h, t_o + kh)$, for k = 1, contains the data minimum, and is bounded.)
2. Partition of shapes according to mean, variance and joint consistency.
3. Partition of shape level sets according to skewness. This is almost the same as the shape level sets, except that zero skewness will include all of the shape level sets for symmetric shapes, for skewness = zero.
4. Partition of an interval of bin width values according to Simonoff-Udina shape stability criterion.
5. Partition of shape level sets according to number of bins, or a range of numbers of bins.
6. Partition or selection of shape level sets according to a specific bin width or range of bin widths.
7. Partition $\{(t_o,h)\}$ into (bounded) $\boldsymbol{D}_o(k)$ according to index of bin, $[t_o + (k–1)h, t_o + kh)$, containing data minimum.
8. Select confidence sets of shapes according to skewness, MISE, maximum likelihood.